# Short Circuit and Arc Flash Study on a Microgrid Facility

Konrad Erich Kork Schmitt
Global Laboratory for Energy Asset
Management and Manufacturing
Texas Tech University
2500 Broadway, 79409
Lubbock, TX, USA
konradkorkschmitt@ieee.org

Cesar A. Negri
Wind and Science
Engineering Department
Texas Tech University
2500 Broadway, 79409
Lubbock, TX, USA
Cesar.Negri@ttu.edu

Saeed Daneshvardehnavi
Electrical and Computer Engineering
Department
Texas Tech University
2500 Broadway, 79409
Lubbock, TX, USA
Saeed.Daneshvardehnavi@ttu.edu

Stephen Bayne, Ph.D.
Electrical and Computer Engineering Department
Texas Tech University
2500 Broadway, 79409
Lubbock, TX, USA
Stephen.Bayne@ttu.edu

Michael Giesselmann, Ph.D.
Electrical and Computer Engineering Department
Texas Tech University
2500 Broadway, 79409
Lubbock, TX, USA
Michael.Giesselmann@ttu.edu

*Abstract*—Arc flash is one of the main hazards when operating an electrical facility. Without correct Personal Protective Equipment (PPE) supplemented by proper design of the protective relays, the operator can be subjected to severe including fatal injuries. By code, facilities are required to properly label their electrical equipment that may be accessed by any operator. While energized, the operator's proximity to the equipment can provide the necessary potential for an incident arc flash accident. The labels are mainly responsible to display the equipment's short circuit and arc flash levels, as well as the minimum PPE level required to operate it. These electrical hazard aspects become more critical in testbed facilities, usually located inside research centers and universities, where the electrical equipment is more easily and frequently accessed by students and researchers. This paper develops complete modeling of a real microgrid testbed facility to perform short circuit and arc flash studies with the main goal to label the devices accessed by the facility's researchers.

## I. Introduction

According to the Occupational Safety and Health Administration (OSHA), from the United States Department of Labor, in 2019, 106 work-related cases of electrical accidents were investigated by OSHA. 59.4% of the accidents resulted in a fatality [1]. Electrical systems can provide both direct and indirect risks. The main danger is related to what the electrical current can do to a human body, such as stoppage of breathing or burns. Associated indirect dangers are the damages to the human body that result from a fall, an explosion, or a fire [2]. An electric shock occurs with the contact of a human body to any voltage source able to provide sufficient current flow through the person's muscles or nerves. The human body can feel currents from in the milliampere range and, depending on the current path through the body the current level can electrocute a person [2]-[3].

An arc flash is an explosion releasing heat, hot gases, and molten metal caused by a short circuit of energized conductors. An arc flash can produce by the burning of the operator's clothing and skin. Injury to bare skin begins at an incident energy value of about 1.2 Cal/cm². This amount of heat energy is equivalent to exposing the skin to a candle flame for about 1 second [4]-[5]. To avoid any of these injuries, operators must wear Personal Protective to energize the systems Equipment (PPE). The PPE is a composite of different items that provide electrical isolation and safety to the operator when the person is close to electrical equipment. There are four different categories of PPE whose protection levels increase according to the incident energy available in the electrical equipment [6]. However, even proper PPE will not guarantee safety. it just assures that second-degree skin burns will not happen [7].

By considering the standards, facilities need to provide short circuits, incident energy, PPE level, and other relevant information for each electrical equipment that may want examination, adjustment, service, and maintenance. It may create a potential for arc flash incident occurrence during energizing.

This information must be organized based on the label and displayed in a despicable place in front of the equipment, keeping the operator aware of the potential for injury associated with the installation. The labels should be installed on all existing enclosure doors and removable panels that the operators may access to achieve a maximum level of safety inside the facility. Once that the facility decides to have an electrical safety program, it is required to pass through a sequence of steps and calculations to understand the hazard level that their energized equipment can provide to the operators.

The main steps to perform in an arc flash study are:

*1) Data collection:* collect and organize the equipment and system's parameters and information.

*2) System's topologies of operations:* define the different configurations in which the system can operate.

*3) Three-phase bolted fault currents:* calculate the short-circuit current and its contribution to arc flash for each piece of equipment.

*4) Arcing current and incident energy:* obtain the arcing current and incident energy through the standard guides based on the system's characteristics.

*5) PPE Category:* Selecting the minimum PPE level required to operate each piece of equipment.

*6) Equipment labeling:* display all the final information in a label for each piece of equipment, pasting it to a place of easy view.

Microgrids are turning to be an indispensable part of modern electrical systems. They operate in both islanding and grid-connected modes [8],[9]. The importance of microgrids has resulted in a significant amount of related research. Different aspects such as power quality [8]-[10], islanding detection, integration of renewable resources (wind and solar), and integration of other assets such as electric vehicles and electrical machines and generators are reported in [11]-[14]. Different research approaches focus on simulation and hardware testing of microgrids. Using a Real-Time Digital Simulator is a reliable option especially for initial studies that are used besides the real testbed [15]-[18]. Beyond that, a real microgrid can assist researchers to do more research. Using previous data to predict the behavior of MG during and after faults is an interesting topic for future research. For instance, [] are two papers in other application which can be applied to arc flash.

Texas Tech University owns and operates the Global Laboratory for Energy Asset Management and Manufacturing (GLEAMM) microgrid testbed facility. The GLEAMM microgrid is located in Lubbock, TX, USA, and is a research facility of Texas Tech University that encourages partnerships between academy and industry, having daily students and researchers working on its equipment [19].

This paper explains the complete short circuit and arc flash study developed for the GLEAMM microgrid. The remaining part of this paper is subdivided as follows. Section II shows the facility's parameters and operation configurations. Section III develops a complete short circuit study for each piece of equipment. Section IV provides the arcing current and incident energy calculations, based on the standard guide. In section V, all the calculation results and PPE dimensioning are shown, as well as the required labeling for each device. At last, section VI concludes the study accomplished in this paper, emphasizing the importance of an arch flash study under a real testbed facility and presenting the future steps for this study.

## II. MICROGRID TESTBED TOPOLOGY

The GLEAMM laboratory focuses on advancing new university innovations and certifying next-generation industry technologies for protecting, enhancing, and managing electricity transmission, and distribution systems. The facility's main goal is to provide an MW-scale testbed for advanced studies in Microgrid areas related to grid modernization, such as energy management, power quality, control, and operation. Figure 1 shows an overview of the facility's diagram and the devices' connections.

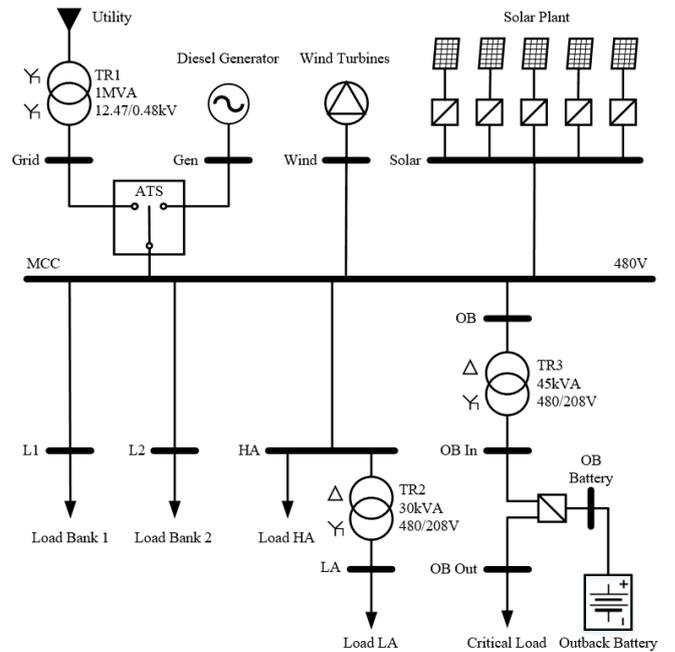

Fig. 1. Microgrid's One-line Diagram Topology.

The microgrid facility feeds critical and non-critical loads that are fed by the following assortment of power sources:

*1) Utility:* The system is mainly connected to the utility's 12.47kV feeder.

*2) Diesel Generator:* It is rated at 569kVA and works as a backup source in case the grid is not available.

*3) Wind Turbines:* Three wind turbines are connected to the microgrid main bus, limited to 400 kVA power generation.

*4) Solar Plant:* The PV system is comprised of five parallel inverters fed by 5 groups of PV panels with a total peak power rating of 150kW.

*5) Dynamic Load Banks:* The facility has two resistive load banks rated in 500kW each and one inductive load bank rated in 187.5kVAr.

*6) Basic Loads* Are the loads responsible for the building's lighting, air conditioning, and general devices.

*7) Outback System:* This is comprised of one hybrid inverter bank and a lead-acid battery of 30kW, being mainly responsible for the critical loads.

Three power transformers are connected to the main microgrid bus, the *MCC*. *TR1* is responsible for step down the utility's 12.47kV to 480V, the most used voltage level in the facility. Moreover, the *MCC* has two 480/208V step-down transformers, *TR2,* and *TR3*. These two devices are responsible to connect some of the non-critical loads and the Outback system in the main bus. Table I displays the three transformers' parameters.

TABLE I. TRANSFORMERS PARAMETERS

| ID | Parameters | | | | |
|---|---|---|---|---|---|
| | S [kVA] | Vp [kV] | Vs [kV] | IZ [%] | X/R |
| TR1 | 1000 | 12.47 | 0.480 | 6.4 | 7 |
| TR2 | 30 | 0.48 | 0.208 | 3.91 | 0.76 |
| TR3 | 45 | 0.48 | 0.208 | 4.07 | 1.26 |

All the devices, except for the Outback inverter bank and HA panel, are located outside the microgrid's building, being connected by underground cables to the MCC bus. The cable connections were modeled o keep the circuit's fidelity based on their distance, type, and several sets by phase. Table II provides the cable parameters for all the main lines connected in the MCC bus. , the reactance values of the cables were neglected for the presence modeling.

TABLE II. LINE PARAMETERS

| Bus | | Parameters | | |
|---|---|---|---|---|
| From | To | Distance [m] | Cable Type | Impedance [Ohm] |
| Grid | MCC | 50 | 500kcmil | 0.00067485 |
| Gen | MCC | 25 | 500kcmil | 0.00101227 |
| Wind | MCC | 300 | 500kcmil | 0.00404910 |
| Solar | MCC | 50 | 500kcmil | 0.00202455 |
| Outback | MCC | 10 | 2AWG | 0.00512795 |
| HA | MCC | 10 | 2AWG | 0.00512795 |
| L1 | MCC | 25 | 500kcmil | 0.00101227 |
| L2 | MCC | 25 | 500kcmil | 0.00101227 |

Based on the system's diagram shown in Fig. 1, the microgrid has two operational topologies. The first topology is with the Automatic Transfer Switch (ATS) connected to the grid. Under this configuration, the facility is directly connected to the utility's feeder, injecting the excess of power to the grid or consuming power to feed its loads when needed. The second possible topology is with the *ATS* connected to the diesel generator. In this situation, the microgrid works in island mode, where the generator is the system's slack bus and the remaining inverter-based sources work as grid-following.

The transition between one topology to another can be planned or not. During a planned transition, the *MCC* bus and its loads are not de-energized. Otherwise, during a non-planned transition, usually due to a grid fault, the *MCC* stays de-energized from the fault until the moment that the generator is turned on and reaches its steady state. During this period, the Outback system is responsible to keep the critical load on, isolating its circuit, and discharging the battery only to support its critical load.

III. SHORT CIRCUIT STUDY

One of the most critical steps in an arc flash study is to estimate the system's short circuit levels. The lack of accuracy and rounding errors in the devices' parameters implies a mismatch between theoretical and practical short circuit levels that the system can provide. Microgrid systems typically have different energy sources and are known to have higher dynamics when compared to power systems grids. Generators, inverters, and intermittent sources increase the system's complexity from the transient state point of view [20].

The GLEAMM microgrid has five energy sources, three of them are inverter-based, namely batteries, solar, and wind plants. The data for short circuit current levels extracted from the datasheets for all the devices. The local utility was able to provide the maximum contribution from the grid in a short condition on the high and low voltage side of *TR1*. In the datasheets of the inverter-based sources, the maximum current that the power electronics devices can provide under a short circuit condition is provided. As shown in Table III, a factor of 1.4 was applied to the inverter-based sources' maximum current values. This factor works as a safety band for any non-standard operation of power electronics devices.

TABLE III. SOURCES PARAMETERS

| Source | Parameters | | | |
|---|---|---|---|---|
| | V [V] | I Max [A] | Factor | Isc Max [A] |
| Utility | 480 | 22131.3 | 1 | 22,130 |
| Generator | 480 | 220.6 | 1 | 220 |
| Wind Turbines | 480 | 660 | 1.4 | 925 |
| Solar | 480 | 181 | 1.4 | 250 |
| Outback Battery [a] | 208 | 169.8 | 1.4 | 240 |

[a.] Current values are based on the bus nominal voltage level, 208V.

Based on each source's maximum contribution, the microgrid is modeled in the PowerWorld software. PowerWorld is a power systems analysis platform widely used by researchers and utilities around the world. With this tool, it is possible to model the system diagram and perform short circuit simulations for different buses and under different operation conditions. In the present study, each source was designed and validated independently. The modeling looked to develop a source that can provide the theoretical maximum short circuit current when a three-phase bolted fault was applied in the source's bus. Figure 2 shows the GLEAMM microgrid modeled on PowerWorld.

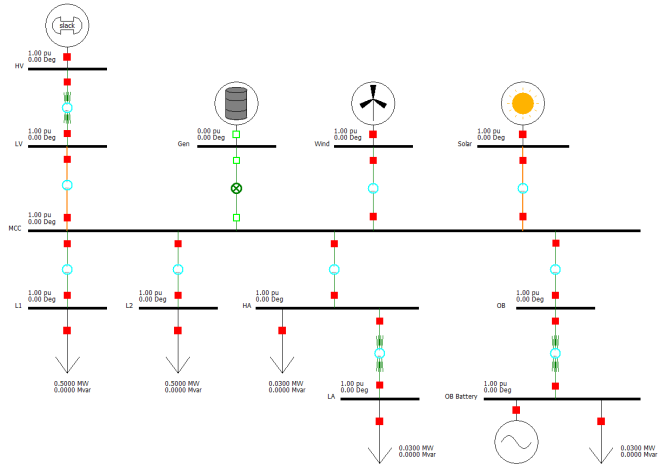

Fig. 2. Microgrid Modeling on *PowerWorld*.

The system was modeled and the maximum current from each source was validated, the short circuit studies were performed. The study aimed to analyze the most relevant buses in the GLEAMM microgrid, as the MCC, Gen, Solar, Wind, OB, HA, L1, and L2. For each of these buses, two three-phase bolted faults were studied, one under the first operation topology, microgrid connected to the utility, and another under the second topology, where the microgrid is connected to the diesel generator. In total, the short circuit study collected twenty three-phase symmetrical fault values, but only the highest value of each bus was used in the arc flash study. Table IV shows the short-circuit study results.

TABLE IV. SHORT CIRCUIT RESULTS

| Bus | | Topology | Isc at Bus [A] | |
|---|---|---|---|---|
| Tag | Voltage [V] | | 1ph SC | 3ph SC |
| MCC | 480 | Grid-Connected | 16147 | 22793 |
| MCC | 480 | Islanded | 1941 | 1296 |
| Generator | 480 | Grid-Connected | 16140 | 22783 |
| Generator | 480 | Islanded | 1941 | 1296 |
| Solar | 480 | Grid-Connected | 16141 | 22773 |
| Solar | 480 | Islanded | 1941 | 1296 |
| Wind | 480 | Grid-Connected | 16121 | 22757 |
| Wind | 480 | Islanded | 1940 | 1296 |
| OutBack | 480 | Grid-Connected | 16111 | 22741 |
| OutBack | 480 | Islanded | 1940 | 1295 |
| OutBack[a] | 208 | Grid-Connected | 8 | 52386 |
| OutBack[a] | 208 | Islanded | 8 | 2989 |
| HA | 480 | Grid-Connected | 16111 | 22741 |
| HA | 480 | Islanded | 1940 | 1295 |
| LA | 208 | Grid-Connected | 8 | 52365 |
| LA | 208 | Islanded | 8 | 2989 |
| L1 | 480 | Grid-Connected | 16140 | 22783 |
| L1 | 480 | Islanded | 1941 | 1296 |
| L2 | 480 | Grid-Connected | 16140 | 22783 |
| L2 | 480 | Islanded | 1941 | 1296 |

a. Current values based on the bus nominal voltage level, 208V.

IV. ARC FLASH STUDY

After the system's data was collected and the maximum short circuit current levels were calculated for each circuit's bus, the following step is to perform the arc flash calculations. There are two main standards to guide arc flash studies in the AC system, the IEEE SA 1584 Guide for Performing Arc-Flash Hazard Calculation [5] and the NFPA 70E Standard for Electrical Safety in the Workplace [6]. To analyze DC systems, Doan's "maximum power method" is suggested, it is used for linear DC systems, but cannot be used properly for solar PV systems where we have a nonlinear current-voltage I-V characteristic [21]. Also, the relevant hazard is considered on the AC side of the solar inverters. The present study was based on and followed the guidance provided by the IEEE SA 1584-2018. As a reference, the NFPA 70E equation does provide results close to the ones obtained in this study.

The main goal of performing an arc flash study is to find the incident energy and the arc-flash boundary levels. The arc flash incident energy represents the amount of thermal energy generated during an electric arc event and the arc flash boundary means the distance from a prospective arc source at which the incident energy is calculated to be 5 J/cm² [5].

The IEEE SA 1584-2018 provides a sequence of equations to find the arc flash parameters, based on short circuit levels and system characteristics. This study is highly related to the facility's setup, mainly with the Electrode Configuration (E.C.), which can be categorized as:

*1) VCB:* vertical conductors/electrodes inside a metal box/enclosure.

*2) VCBB:* vertical conductors/electrodes terminated in an insulating barrier inside a metal box/ enclosure.

*3) VOA:* vertical conductors/electrodes in the open air.

*4) HCB:* horizontal conductors/electrodes inside a metal box/enclosure.

*5) HOA:* horizontal conductors/electrodes in the open air.

The GLEAMM microgrid is only contained E.C. type VCB. This category informs the choice of the E.C. coefficients. the arc duration time (T), the gap distance between conductors (G), the working distance (D), and the enclosure equivalent size (ESS) are other important parameters that must be considered.

The most conservative way to consider the arc duration is to assume the same arcing current on the faulted bus over the fault time, which comes from the slowest tripping device [22]. In this study, the arc duration was considered as 5 cycles. The gap distance was considered as 32 mm and the working distance as 457.2 mm as respectively suggested by Tables 8 and 10 from [5]. Using Table 6 from [5] were obtained an ESS value of 19.999 mm for all of GLEAMM's enclosures, which have a height and width of 508 mm. The ESS parameter enables us to calculate the enclosure size correction factor (CF) by equation (1).

$$CF = b_1 \cdot ESS^2 + b_2 \cdot ESS + b_3 \quad (1)$$

where,

| $CF$ | enclosure size correction factor; |
| $ESS$ | equivalent enclosure size (mm); |
| $b$ | coefficients for a typical VCB E.C., Table 7 of [5]. |

The arc flash study is categorized by systems with voltage levels less or equal to 600V and greater than 600V. The microgrid's main voltage level is 480V, so the study followed the first option. Before getting the average root mean square (RMS) arcing current for our 480V system, we must obtain the arcing current at 600V open circuit voltage through equation (2).

$$I_{arc\_600} = 10^{(k_1 + k_2 \log I_{bf} + k_3 \log G)} \cdot$$
$$(k_4 I_{bf}^6 + k_5 I_{bf}^5 + k_6 I_{bf}^4 + k_7 I_{bf}^3 + k_8 I_{bf}^2 + k_9 I_{bf} + k_{10}) \quad (2)$$

where,

| $I_{arc\_600}$ | average RMS arcing current at $V_{oc} = 0.6\ kV$ (kA); |
| $I_{bf}$ | bolted fault current three-phase fault (kA); |
| $G$ | the gap distance between electrodes (mm); |
| $k$ | coefficients for VCB E.C., Table 1 of [5]. |

Based on the arcing current at 600V open circuit voltage we can obtain the final arcing current for our 480V system with equation (3).

$$I_{arc} = \frac{1}{\sqrt{\left[\frac{0.6}{V_{oc}}\right]^2 \cdot \left[\frac{1}{I_{arc\_600}^2} - \left(\frac{0.6^2 - V_{oc}^2}{0.6^2 \cdot I_{bf}^2}\right)\right]}} \quad (3)$$

where,

$I_{arc}$      final rms arcing current at $V_{oc} = 0.48\ kV$ (kA);

$V_{oc}$      open-circuit voltage $V_{oc} = 0.48\ kV$ (kV).

The arcing current is one of the characteristic parameters that represent the system's hazards. The IEEE SA 1584-2018 provides different equations according to the system voltage level to find the incident energy. As the microgrid is a 480V system, again it is used the equation for any open circuit voltage less or equal to 600V. Based on it, we can obtain the incident energy through equation (4).

$$E_{\leq 600} = \frac{12.552}{50} T \cdot 10^{\left( k_1 + k_2 \log G + \frac{k_3 I_{arc\_600}}{\left(k_4 I_{bf}^7 + k_5 I_{bf}^6 + k_6 I_{bf}^5 + k_7 I_{bf}^4 + k_8 I_{bf}^3 + k_9 I_{bf}^2 + k_{10} I_{bf}\right)} + k_{11} \log I_{bf} + k_{12} \log D + k_{13} \log I_{arc} + \log\left(\frac{1}{CF}\right) \right)} \quad (4)$$

where,

$E_{\leq 600}$      incident energy for $V_{oc} \leq 0.6\ kV$ (J/cm²);
$T$      arc duration (ms);
$D$      working distance (mm);
$k$      coefficients for VCB E.C., Table 3 of [5].

The incident energy level represents the amount of thermal energy that that bus can provide during a short circuit event. This value determines the minimum PPE category required to work in that device when energized. Finally, with the parameters collected and calculated, it is possible to obtain the arc-flash boundary through equation (5). Similarly, to the incident energy equation, the arc flash boundary has different equations for different voltage ranges. In the microgrid's case, with a 480 level, it must use the equation for any open circuit voltage less or equal to 600V.

$$AFB_{\leq 600} = 10^{\left[\frac{\left(k_1 + k_2 \log G + \frac{k_3 I_{arc\_600}}{\left(k_4 I_{bf}^7 + k_5 I_{bf}^6 + k_6 I_{bf}^5 + k_7 I_{bf}^4 + k_8 I_{bf}^3 + k_9 I_{bf}^2 + k_{10} I_{bf}\right)} + k_{11} \log I_{bf} + k_{13} \log I_{arc} + \log\left(\frac{1}{CF}\right) - \log\left(\frac{20}{T}\right)\right)}{-k_{12}}\right]} \quad (5)$$

where,

$AFB_{\leq 600}$      arc-flash boundary for $V_{oc} \leq 0.6\ kV$ (mm);
$k$      coefficients for VCB E.C., Table 3 of [5].

These steps must be followed for each of the system's devices, applying the five equations, and obtaining values that represent the hazards in each bus. Section V compiles and displays all the short circuit and arc flash study results, looking to organize this information in labels that must be paste into each device as a warning of possible hazards.

V. RESULTS AND EQUIPMENT LABELING

Based on the short circuit and arc flash results we can understand the hazards present in each device that the operator can be in touch with. Moreover, with these results we can rate the minimum PPE category required to work on these devices, avoiding possible injuries. The PPE has four categories that depend on the incident energy level [6].

*1) Category 1:* minimum arc rating of clothing 4 cal/cm².

*2) Category 2:* minimum arc rating of clothing 8 cal/cm².

*3) Category 3:* minimum arc rating of clothing 25 cal/cm².

*4) Category 4:* minimum arc rating of clothing 40 cal/cm².

In the GLEAMM's microgrid facility arc flash study, all the buses presented incident energy smaller than 4 cal/cm², assuming then category 1. As the first PPE level, category 1 requires that the operator use an arc-rated long sleeve shirt and pants, as well as a face shield. According to the device's location and access conditions, the operation can also be required to wear an arc-rated jacket, rainwear, parka, and hardhat liner. In addition to these items, the person is also encouraged to wear heavy-duty leather gloves, hardhat, eye protection, hearing protection, and leather footwear.

Table V provides an overview of the performed study results, displaying the bolted three-phase fault current, the arcing current, incident energy, arc-flash boundary and the PPE category for each bus analyzed.

TABLE V. SHORT CIRCUIT AND ARC FLASH RESULTS

| Bus | $V_{oc}$ [V] | $I_{bf}$ [kA] | $I_{arc}$ [kA] | $E_{\leq 600}$ [cal/cm²] | AFB [cm] | PPE Hazard Category |
|---|---|---|---|---|---|---|
| MCC | 480 | 22.79 | 16.81 | 3.6 | 91 | 1 |
| Generator | 480 | 22.78 | 16.85 | 3.6 | 91 | 1 |
| Solar | 480 | 22.77 | 16.85 | 3.6 | 91 | 1 |
| Wind | 480 | 22.76 | 16.82 | 3.6 | 91 | 1 |
| Outback | 480 | 22.74 | 18.82 | 3.6 | 91 | 1 |
| Outback[a] | 208 | 52.38 | 16.12 | 3.8 | 94 | 1 |
| HA | 480 | 22.74 | 16.82 | 3.6 | 91 | 1 |
| LA[a] | 208 | 52.36 | 16.12 | 3.8 | 94 | 1 |
| L1 | 480 | 22.78 | 16.85 | 3.6 | 91 | 1 |
| L2 | 480 | 22.78 | 16.85 | 3.6 | 91 | 1 |

a. Current values based on the bus nominal voltage level, 208V.

With all studies completed, the last step is to label all the devices to easily inform the users about the hazards present inside that encloser. In the facility, we used the "Arc Flash and Shock Hazard" warning label. Figure 3 shows the warning label developed for the GLEAMM's microgrid *MCC* bus.

**WARNING**

**Arc Flash and Shock Hazard**

| | | | |
|---|---|---|---|
| Nominal System Voltage | 480 V | Incident Energy (cal/cm²) | 3.6 |
| Arc Flash Boundary | 0.91 m | Working Distance | 0.46 m |
| Limited Approach | 1 m | OR | |
| Restricted Approach | 0.3 m | PPE Hazard Category | 1 |
| | | Arc Rating of Clothing | 4 cal/cm² |

Arc-rated PPE: ☒ Face shield ☐ Coverall  Additional PPE: ☐ Hard hat  ☒ Leather footwear
☒ Long-sleeve shirt ☐ Balaclava ☐ Hard hat liner  ☐ Safety goggles
☐ Flash suit jacket ☒ Gloves  ☒ Safety glasses
☐ Flash suit pants ☐ Jacket  ☒ Hearing protection
☐ Flash suit hood ☐ Parka  ☒ Heavy duty leather gloves
☒ Pants ☐ Rainwear

Equipment ID: MCC

Fig. 3. MCC Bus Labeling.

This label displays the system nominal voltage, the arc-flash boundary, incident energy, working distance, PPE category, and the required and additional PPE clothes for work if the respective device is energized. Moreover, with the NFPA 70E, we have a limited and restricted approach distance that is based on the system's nominal voltage. All this information is neatly on the label and must be pasted in an easy view place.

## VI. FINAL CONSIDERATIONS

The present paper proposed to develop a complete short circuit and arc flash study for a microgrid facility, the GLEAMM installation. As a research center, the GLEAMM microgrid daily has students and researchers on its visiting the facility that may be close and even operating some of the de-energized and energized devices. The microgrid presented a considerable short circuit level in all of its buses, 22.75 kA on average with five different power sources. The arc flash study looked to measure the hazards present in the facility, obtaining the incident energy and the arc-flash boundary level for each of the system's buses. The purpose of calculating these parameters is to inform the operators of the hazards that they can face inside the facility. The importance of this study is emphasized when considered that incident energy of 1.2 cal/cm² is enough to injury a bare skin and the microgrid's average incident energy was 3.63 cal/cm². Therefore, even that the GLEAMM microgrid has the function to be a real testbed, open for new researchers and looking to improve the students' practical skills, all persons, when working with the facility, must be aware of its hazards and understand that without the right PPE category that place has the potential to provide them different types of injuries. The presented study will take continuity analyzing the relation between the different generation levels that each resource has with the short-circuit and incident energy levels. GLEAMM has a historical database with generation levels measured in the range of seconds. The next research will collect the data and with a probabilistic analysis will check which are the most common combination between solar and wind generation and their arc-flash levels. As well as with the statistical analysis, a short circuit and incident energy daily profile will be computed, showing which is the safer and most reasonable time to make maintenance based on the short circuit capability of the facility along its historical operation day.

ACKNOWLEDGMENT

The authors would like to acknowledge the Texas Tech University, the Reese Technology Center, and the Global Laboratory for Energy Asset Management and Manufacturing (GLEAMM) for all the structural and financial support to develop the present research.